\documentclass[12pt]{iopart}

\usepackage{graphicx}
\usepackage{hyperref}
\begin{document}

\title[Quarkonium in ALICE]{Quarkonium production measurements with the ALICE detector at the LHC}

\author{Gin\'es Mart\'{\i}nez Garc\'{\i}a, for the ALICE collaboration}

\address{Subatech, CNRS/IN2P3, LUNAM Universit\'e, Ecoles de Mines de Nantes, Universit\'e de Nantes, Nantes, France}
\ead{gines.martinez@subatech.in2p3.fr}
\begin{abstract}
 In this new energy regime, quarkonium provides a unique probe to study the properties of the high-density, strongly interacting system formed in the early stages of high-energy heavy-ion collisions. In ALICE, quarkonium states are reconstructed down to $p_{\rm T}$ = 0 via their $\mu^+\mu^-$ decay channel in the muon spectrometer ($2.5\leq y \leq 4.0$) and via their e$^+$e$^-$ channel in the central barrel ($|y|\leq 0.9$). Measurement of the transverse momentum and rapidity distributions of inclusive J/$\psi$ production cross section in proton-proton collisions at $\sqrt{s}$=2.76 and 7 TeV will be presented. We will discuss the dependence on charged particle multiplicity of the inclusive J/$\psi$ yield in proton-proton collisions at $\sqrt{s}$ = 7 TeV. Finally, the analysis of the inclusive J/$\psi$ production in Pb-Pb collisions at $\sqrt{s_{\rm NN}}$ = 2.76 TeV will be described. Preliminary results on the nuclear modification factor ($R_{\rm AA}$) and the central to peripheral nuclear modification factor ($R_{\rm CP}$) will be discussed. 
\end{abstract}

\section{Quarkonium in heavy ion collisions}
Quarkonium was proposed as a probe of the QCD matter formed in relativistic heavy-ion collisions more than two decades ago. 
A familiar prediction, quarkonium suppression due to color-screening of the heavy-quark potential in deconfined QCD matter \cite{Satz86}, has been experimentally searched for at the SPS and RHIC heavy-ion facilities. A less familiar prediction was also devised at the same time: quarkonium enhancement due to heavy-quark stopping in deconfined QCD matter \cite{Svet88}.

The NA50 experiment at SPS reported the observation of J/$\psi$ and $\psi$' suppression in central heavy-ion collisions \cite{NA5005}.  
This observation was also reported for proton-induced reactions on different target nuclei, from Be to Pb \cite{NA5006}.
Quarkonium absorption in cold nuclear matter (CNM) has been hypothesized as the mechanism responsible for quarkonium suppression at mid-rapidity in pA collisions.  
A model based on Glauber theory, has been developed to estimate the magnitude of quarkonium suppression in central heavy-ion collisions due to quarkonium absorption in CNM. 
The magnitude of J/$\psi$ suppression due to dissociation in hot QCD matter was then measured to be about 40\% \footnote[7]{which means that 60\% of the initial  produced J/$\psi$ survives the hot QCD matter formed in central Pb-Pb collisions at SPS.}. One has to consider that the measured J/$\psi$ production consists of prompt-J/$\psi$ ($\sim65$\%) and decay-J/$\psi$ from higher resonances like $\chi_{\rm c}$ and $\psi'$. The observed J/$\psi$ suppression would be compatible with the suppression of $\chi_{\rm c}$ and $\psi$' resonances. 
The NA60 experiment measured J/$\psi$ production in pA collisions at the same CM collision energy as the heavy-ion measurements and addressed the dependence of quarkonium suppression with the laboratory proton beam-energy \footnote[3]{J/$\psi$ absorption cross-section was assumed to be independent of the proton beam-energy. It should be noted that the NA60 J/$\psi$ acceptance is $0<y_{\rm CM}<1$, consequently Ap collisions would be needed.}.
A new evaluation of the quarkonium absorption in CNM was performed and a first attempt to consider the parton shadowing was addressed. 
The outcome of this analysis is that only about 20-30\% of  the suppression in the most central Pb-Pb collisions at SPS energies is indeed due to dissociation in hot QCD matter \cite{NA6009}.

The PHENIX experiment at RHIC has reported the observation of J/$\psi$ suppression in central Au-Au collisions at $\sqrt{s_{\rm NN}}$=200 GeV (10 times higher than the maximum energy in the CM at SPS) \cite{PHENIX07}. Deuteron-gold collisions have been utilized to measure CNM effects at RHIC energies \cite{PHENIX10}.  As of today, only parton-shadowing and quarkonium-absorption mechanisms  have been considered and the d-Au J/$\psi$ results are not fully understood. As a consequence, J/$\psi$ suppression due to dissociation in hot QCD matter is roughly estimated to be 40-80\% in central Au-Au collisions at RHIC energies. This result would suggest higher suppression than that observed at SPS.  In addition, the large rapidity coverage of the PHENIX experiment enables measurements at large rapidity, showing that the suppression is larger than that observed at mid-rapidity. This is a very intriguing experimental observation, and it has not yet been understood whether its origin is due predominantly to hot or cold nuclear matter effects. Finally, the STAR experiment has measured a smaller suppression at high transverse momentum ($p_{\rm T}\geq 5$ GeV/$c$) at mid-rapidity \cite{STARQM}  although the experimental errors remain large.

The LHC collider has opened a new energy regime for the study of quarkonium in heavy-ion collisions. At these energies (15 times higher than the maximum energy in the CM at RHIC, and 150 times higher than that at SPS), on average one J/$\psi$ particle is expected to be produced in every central Pb-Pb collision, together with about 50-100 $c$$\bar{c}$ quark pairs. As suggested in 1988 \cite{Svet88}, under these conditions the charm quark yield per unit of rapidity could be large enough to enhance the charmonium production in later phases of the hot QCD-matter dynamical evolution, in particular when the energy density is low enough to enable the charmonium bound state to be formed \cite{PBM00,Andronic11}. At LHC, J/$\psi$ is abundantly produced, allowing for detailed studies of its production, such as azimuthal asymmetry, polarization, and $R_{\rm AA}$\footnote{The nuclear modification factor $R_{\rm AA}$ is defined as the ratio of the yield measured in nucleus-nucleus (AA) to that expected on the basis of the proton-proton yield scaled by
the number of binary nucleon-nucleon collisions in the nucleus-nucleus reaction.} and their dependence on rapidity and transverse momentum. Bottomonium resonances are being studied at RHIC and LHC and are complementary to the charmonium experimental observations at RHIC and at LHC. In particular, the bottom rapidity density at LHC is expected to be similar to that of charm at RHIC, and color-screening of the $\Upsilon$(2S) resonance should be similar to that of the J/$\psi$. The first results on the J/$\psi$ $R_{\rm AA}$ at LHC down to $p_{\rm T}$=0 will be presented in this talk. More detailed description of this analysis can be found in \cite{Pillot11}.

The baseline of quarkonium studies in heavy-ion collisions at LHC, as in the previous studies at lower energies, requires the study of quarkonium production in proton-proton and proton-nucleus collisions. The LHC provides huge proton-proton luminosities and should provide pPb collisions in the future. Quarkonium studies in proton-proton collisions at this new energy regime might allow further insight to understand its production mechanisms. Furthermore, it has been argued that high-multiplicity pp collisions at LHC could lead to the formation of high energy density QCD matter, similar to heavy ions collisions \cite{Werner11}. Indeed, the charged particle multiplicity reached in pp collisions at the LHC is similar to that measured in semi-peripheral Cu-Cu collisions at $\sqrt{s_{\rm NN}}$= 200 GeV \cite{ALICE10a, Phobos11}. J/$\psi$ production results in pp collisions at $\sqrt{s}$= 2.76 TeV and 7 TeV will be presented.  The dependence on charged particle multiplicity (up to five times the mean charged particle density) of the J/$\psi$ yields in pp collisions at $\sqrt{s}$ = 7 TeV will be discussed. More detailed information about these analysis can be found in \cite{Arnaldi11}.

\section{ Experimental apparatus and running conditions}

ALICE \cite{Schukraft11} is a general purpose heavy-ion experiment. It consists of a central part with a pseudo-rapidity coverage of $|\eta|<$0.9, and a muon spectrometer placed at large rapidity ($-4 < \eta< -2.5$). J/$\psi$  production is measured down to $p_{\rm T}$=0 in both rapidity regions, at mid-rapidity in the dielectron and at large rapidity in the dimuon decay channel \cite{ALICE11a}. 

The main detectors used for the dielectron analysis are the Inner Tracking System (ITS) and the Time Projection Chamber (TPC). The ITS, placed at radii between 3.9 cm and 43 cm, consists of six cylindrical layers of silicon detectors, equipped with two layers each of silicon pixel, silicon drift and silicon strip detectors. The large cylindrical TPC has full azimuthal coverage, with a length of 5 m and radial coverage from 85 cm to 247 cm. The TPC provides particle identification (PID) via the measurement of the specific energy loss (${\rm d}E/{\rm d}x$) of particles in the detector gas. Its excellent ${\rm d}E/{\rm d}x$ resolution of 5.5\%, allows for electron ID.  The Time-Of-Flight (TOF) enhances the TPC electron ID in Pb-Pb analysis. The Transition Radiation Detector (TRD) and the Electromagnetic Calorimeter (EMCAL) detectors are not used in the analysis reported here, but are expected to significantly improve the electron identification and triggering capabilities of the experiment in the future. For the dimuon analysis, the muon spectrometer consists of a frontal absorber followed by a 3 T$\cdot$m dipole magnet, coupled to tracking and triggering devices. Muons produced in the $-4<\eta<-2.5$ rapidity region are filtered by means of a 10 interaction length ($\lambda_{\rm I}$) thick frontal absorber made of carbon, concrete and steel, and placed between z=-0.9 and z=-5.0 m from the nominal position of the Interaction Point (IP). Muon tracking is carried out by five tracking stations, positioned between z=-5.2 and z=-14.4 m from the IP, each one based on two planes of Cathode Pad Chambers (CPC). Stations 1 and 2 (4 and 5) are located upstream (downstream) of the dipole magnet, while station 3 is embedded inside the magnet. A muon triggering system is placed downstream of a 1.2 m thick iron wall (7.2 $\lambda_{\rm I}$), which absorbs frontal-absorber punch-through hadrons, secondary hadrons escaping the frontal-absorber and low-momentum muons ($p < 1.5$ GeV/$c$). The muon triggering system consists of two stations positioned at z=-16.1 and z=-17.1 m from the IP, each equipped with two planes of Resistive Plate Chambers (RPC). Throughout its full length, a conical absorber  made of tungsten, lead and steel protects the muon spectrometer against secondary particles produced by the interaction of large-$\eta$ primaries in the beam pipe. The forward VZERO detectors, made of two scintillator arrays, covering the range $2.8 < \eta < 5.1$ and $-3.7 < \eta < -1.7$, and positioned at z=340 and z=-90 cm from the IP respectively, are used for triggering, beam-gas rejection and centrality selection purposes. Finally, the zero degree calorimeters (ZDC) placed 116 m down and up-stream ALICE IP are used for rejecting electromagnetic Pb-Pb interactions and satellite Pb-Pb collisions.

In proton-proton collisions the ALICE minimum bias trigger (MB) is defined as the logical OR between the requirement of at least one hit in the ITS pixel layers (SPD), and a signal in one of the two VZERO detectors. ALICE MB trigger selects about 87\% of the proton-proton inelastic cross section. The cross-sections of the logical AND of the VZERO detectors were measured during a van der Meer scan in pp collisions at 7 TeV and 2.76 TeV \cite{Oyama11} and were used for the absolute normalization of the inclusive J/$\psi$ cross-section.  The muon trigger ($\mu$-MB), which requires, in addition to a MB event, at least one muon in the angular acceptance of the muon spectrometer that fires the muon trigger system, is used to take advantage of the full luminosity delivered by the LHC at the IP. The luminosity enrichment factor of the $\mu$-MB triggered data samples, with respect to the MB trigger, ranges between 1 and 100, depending on the LHC instantaneous luminosity at the ALICE IP, which ranges between 1.6 $\cdot$ (10$^{28}$-10$^{30}$) cm$^{-2}$s$^{-1}$. The logical  AND between the ITS pixel layer and the VZERO detector was used as the  ALICE MB trigger in Pb-Pb collisions. The MB trigger in Pb-Pb collisions selects about 97\% of the Pb-Pb total hadronic cross section. The centrality of the collision is determined from the amplitude of the VZERO signal fitted with a geometrical-Glauber model. The Pb-Pb analysis  has been limited to the 0-80\% most central collisions, where the contamination from electromagnetic interactions is negligible \cite{Toia11}.

\section{Measurement of J/$\psi$ inclusive cross sections  in proton-proton collisions at 2.76 TeV and 7 TeV}
J/$\psi$ production is being studied extensively in ALICE: as a function of the centre of mass energy, transverse momentum, rapidity and of the charged particle density. Furthermore, a polarization analysis is being performed and first performance results of J/$\psi$ from B mesons decay have been obtained.  A total integrated luminosity of about 125 nb$^{-1}$ was collected by ALICE in pp collisions at 7 TeV in 2010 in the $\mu$-MB trigger and about 10 nb$^{-1}$ in the MB trigger. The J/$\psi$ signal is clearly seen in the invariant mass distribution of the dimuon and dielectron pairs.  A mass resolution of $\approx$85 MeV/c$^2$  ($\approx$40 MeV/c$^2$) in the dimuon (dielectron) channel is measured and is well described by the MC detector simulations.
\begin{figure}[h]
   \centering
  \includegraphics[width=7.4cm]{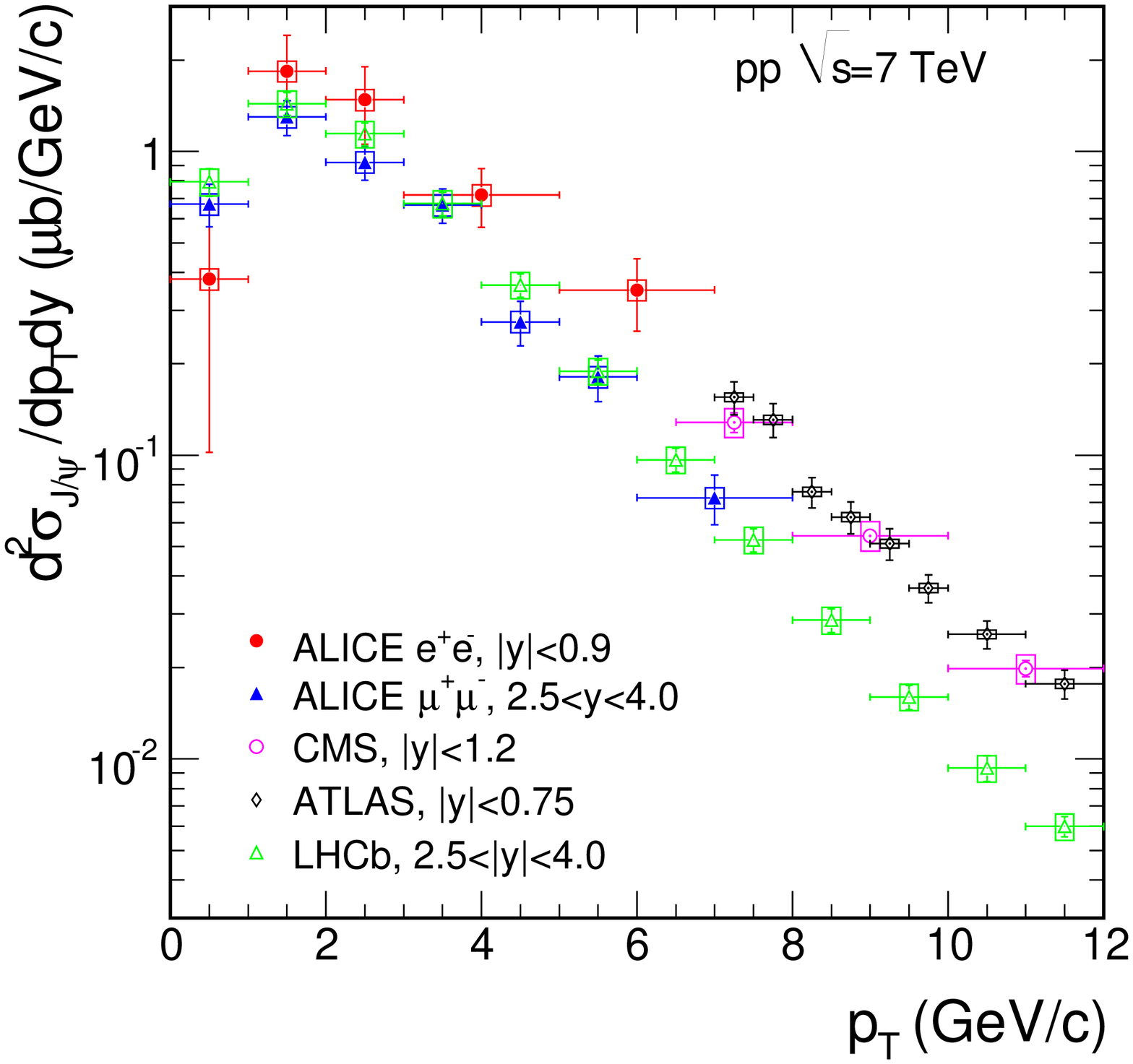} 
  \includegraphics[width=8.0cm]{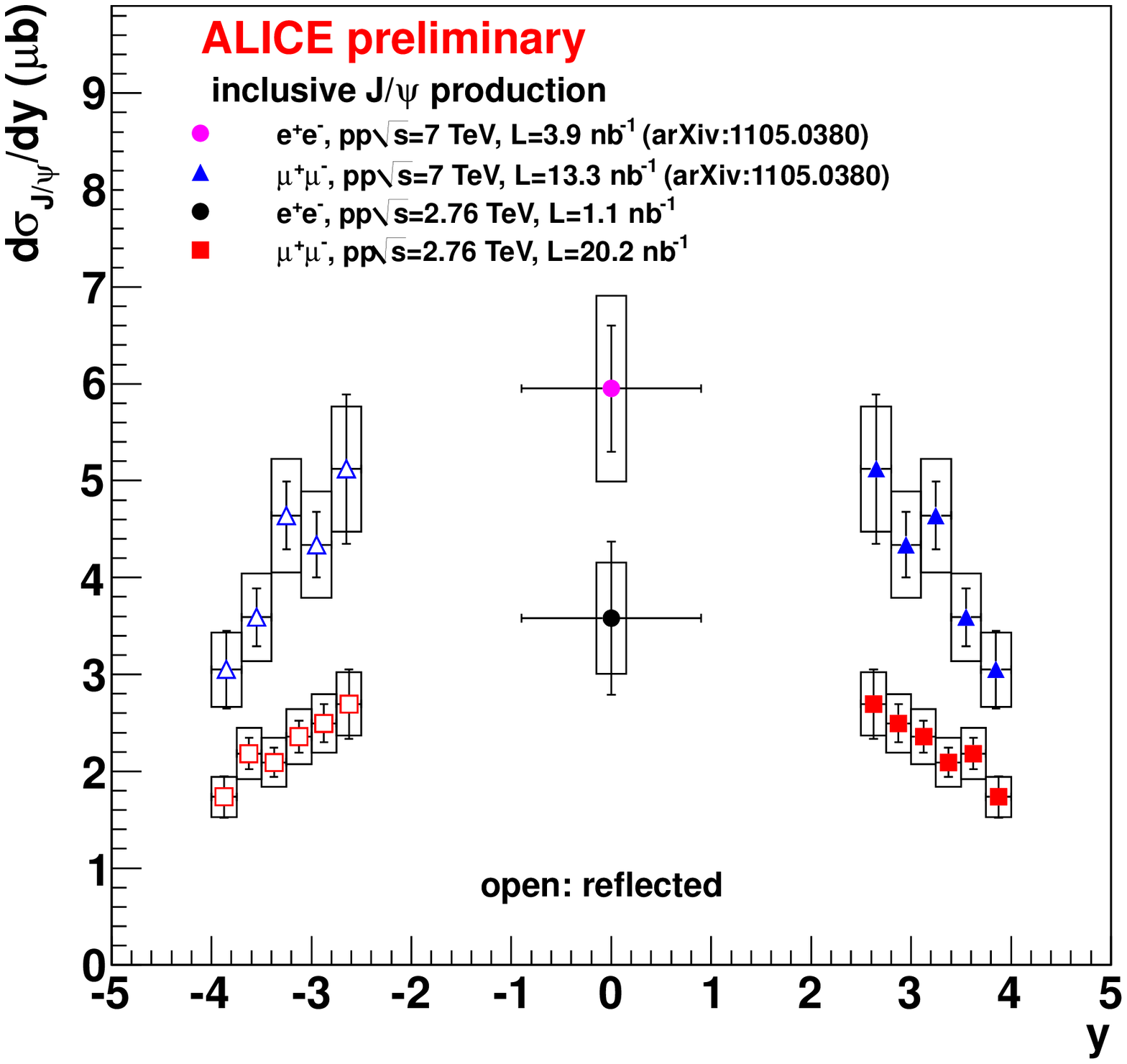}
   \caption{Left: $d^2\sigma_{J/\psi}/dp_{\rm T}dy$ compared with results from the other LHC experiments, obtained in similar rapidity ranges \cite{ALICE11a}.  Right: Inclusive J/$\psi$ rapidity differential cross-section down to $p_{\rm T}$=0 in pp collisions at 2.76 and 7 TeV.}
   \label{fig:JpsiInpp}
\end{figure}
The total inclusive J/$\psi$ cross sections and the differential inclusive cross sections in transverse momentum and rapidity were obtained for a data sample of 15.6 nb$^{-1}$ (3.9 nb$^{-1}$) for the dimuon (dielectron) channel at 7 TeV \cite{ALICE11a}. Good agreement with ATLAS, CMS and LHCb experiments is found in the common ($p_{\rm T}$, y) regions (see Fig.\ref{fig:JpsiInpp}-left).  ALICE has unique coverage at the LHC, measuring the inclusive J/$\psi$ rapidity distribution in a large rapidity range, down to $p_{\rm T}$=0  (see Fig.\ref{fig:JpsiInpp}-right). Details of this analysis can be found in reference \cite{ALICE11a}.
In March 2011, LHC provided pp collisions at $\sqrt{s}$=2.76 TeV, the same energy as for Pb-Pb collisions. This analysis provides essential reference data to measure the nuclear modification factor of the J/$\psi$ production in Pb-Pb collisions. The integrated luminosity for the analysis is $L_{\rm int} = 20.2$ nb$^{-1}$ in the dimuon channel, and  $L_{\rm int} = 1.1$ nb$^{-1}$  in the dielectron channel. The corresponding inclusive J/$\psi$ rapidity differential cross-section down to $p_{\rm T}$=0 is shown in Fig.\ref{fig:JpsiInpp}-right \cite{Arnaldi11}. The phenomenological J/$\psi$ interpolation to 2.76 TeV described in reference \cite{Bossu11} agrees with this analysis.

\begin{figure}[h]
   \centering
  \includegraphics[width=7.7cm, height=6.1cm]{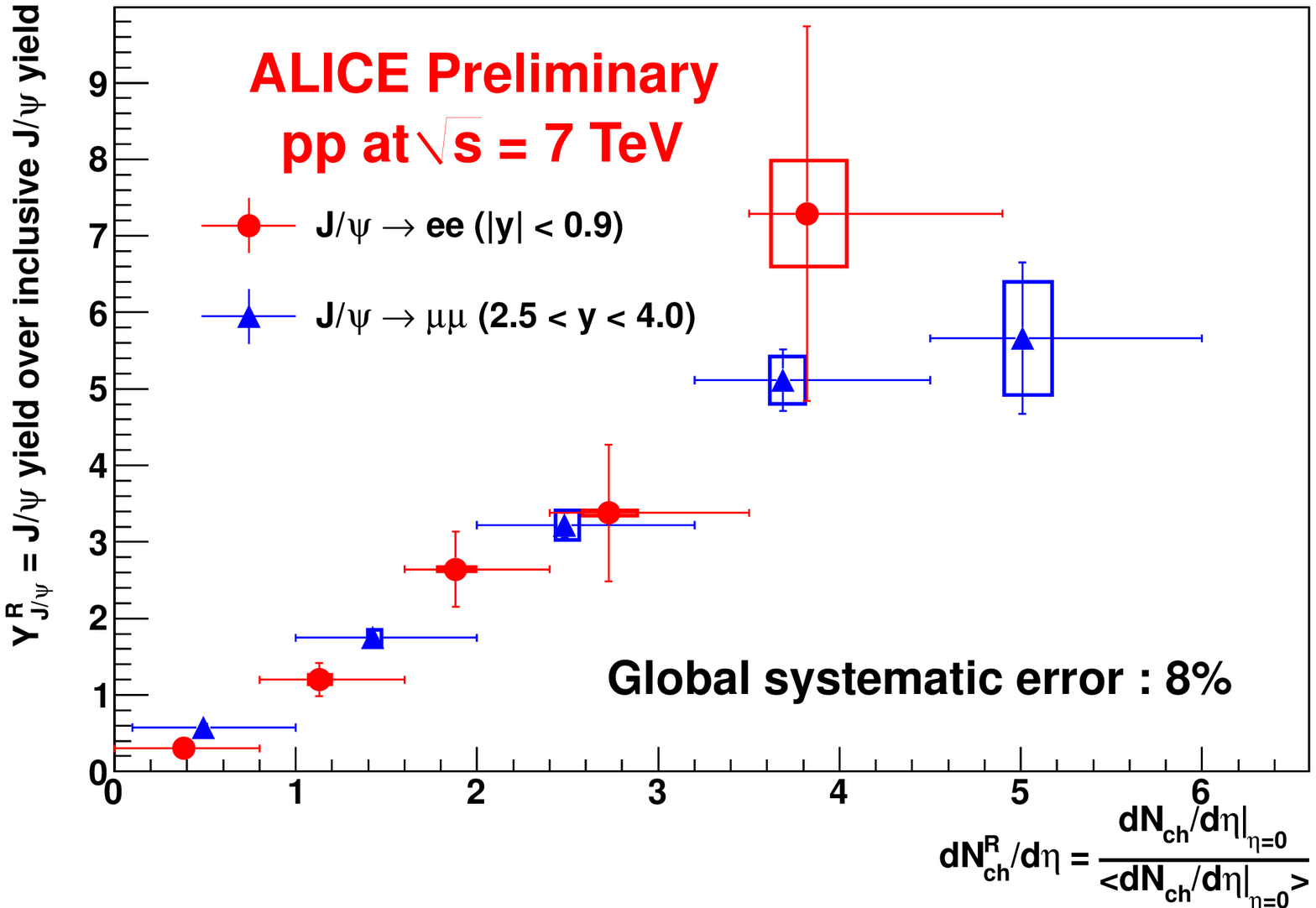} 
  \includegraphics[width=7.7cm, height=6.1cm]{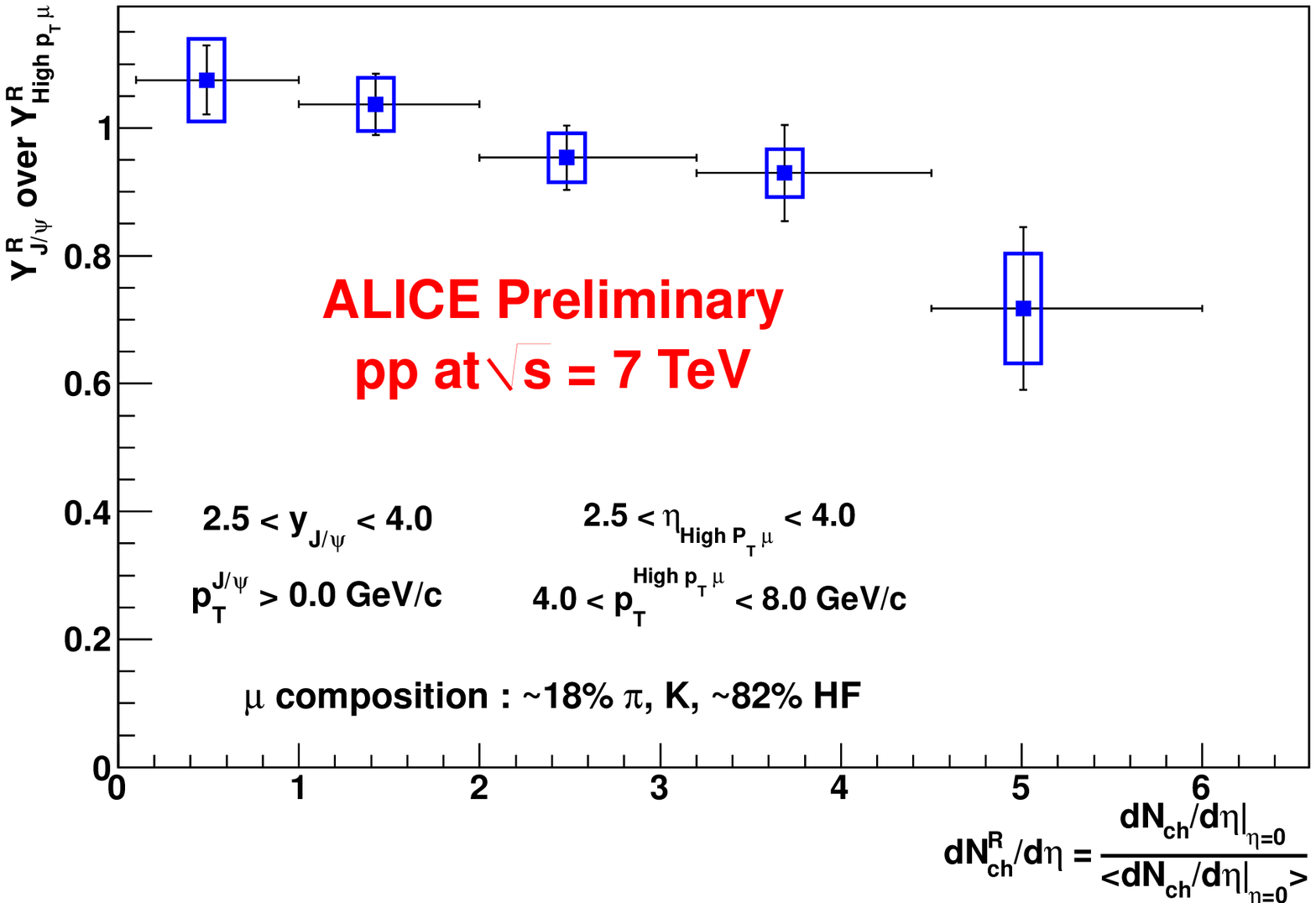}
   \caption{Left: Dependence of the inclusive J/$\psi$ yield in $|y|<0.9$ and $2.5<y<4$ on the charged particle density at mid-rapidity in pp collisions at 7 TeV. Right: The ratio of the J/$\psi$ yield to that of the inclusive muon yield in the $p_{\rm T}$  range $4<p_{\rm T}<8$ GeV/$c$  in the rapidity domain $2.5<y<4$ as a function of $dN_{\rm ch}/d\eta$.}
   \label{fig:JpsiInHighMult}
\end{figure}

ALICE has measured the dependence of the inclusive J/$\psi$ yield on the charged particle density ($dN_{\rm ch}/d\eta$) in pp collisions at 7 TeV. Charged particle densities were measured in the pseudo-rapidity range $|\eta|<1.6$ with the ITS pixel detector. A linear dependence of the J/$\psi$ yields in both rapidity domains on $dN_{\rm ch}/d\eta$ (see Fig.\ref{fig:JpsiInHighMult}-left) is observed. Note the rapidity gap of ~3.25 rapidity units between mid-rapidity measurements (J/$\psi \rightarrow {\rm e}^+{\rm e}^-$ and  $dN_{\rm ch}/d\eta$) and that at large rapidity (J/$\psi \rightarrow \mu^+\mu^-$) has to be noted.
Fig. \ref{fig:JpsiInHighMult}-right panel, shows the ratio of the J/$\psi$ yield to that of the inclusive muon yield in the $p_{\rm T}$  range $4<p_{\rm T}<8$ GeV/$c$ in the same rapidity domain, as a function of $dN_{\rm ch}/d\eta$.
About 82\% of the muons produced in this $p_{\rm T}$ domain arise from the semi-leptonic decay of heavy flavoured hadrons \cite{Zhang11}. The J/$\psi$ yield exhibits a weaker increase with $dN_{\rm ch}/d\eta$ than the one of high $p_{\rm T}$ muons. Several different mechanisms could explain this observation, including kinematical effects, modification of the $p_{\rm T}$ distribution, modification of the bottom to charm ratio, etc. Mechanisms based on final state effects cannot be excluded and, interestingly, the effect is most significant at 5 times $\langle dN_{\rm ch}/d\eta \rangle$.

\section{Inclusive J/$\psi$ nuclear modification factor ($R_{\rm AA}$) in Pb-Pb collisions at $\sqrt{s_{\rm NN}}=2.76$ TeV}
\begin{figure}[h]
   \centering
  \includegraphics[width=7.35cm, height=6.1cm]{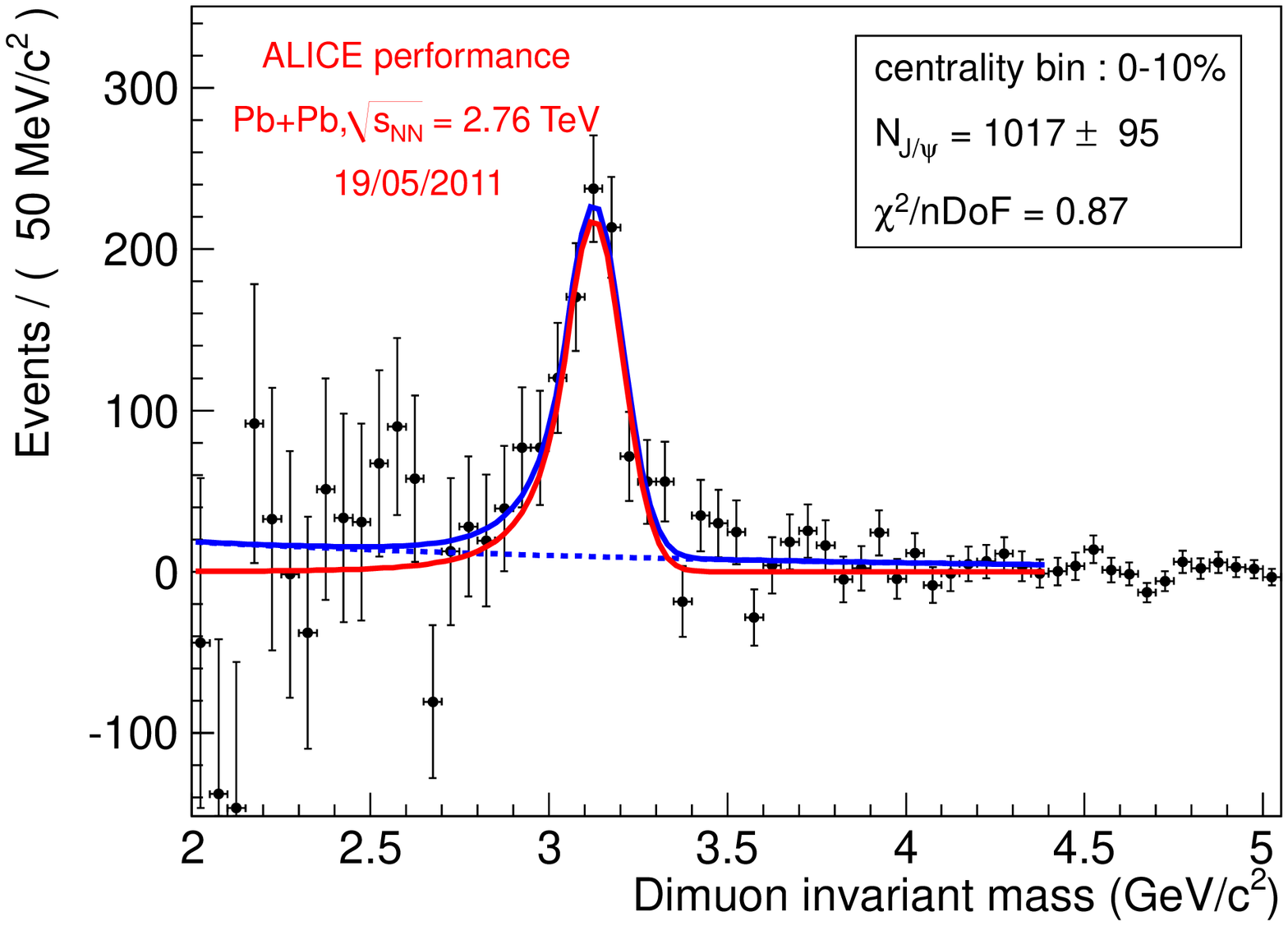} 
  \includegraphics[width=8.05cm, height=6.1cm]{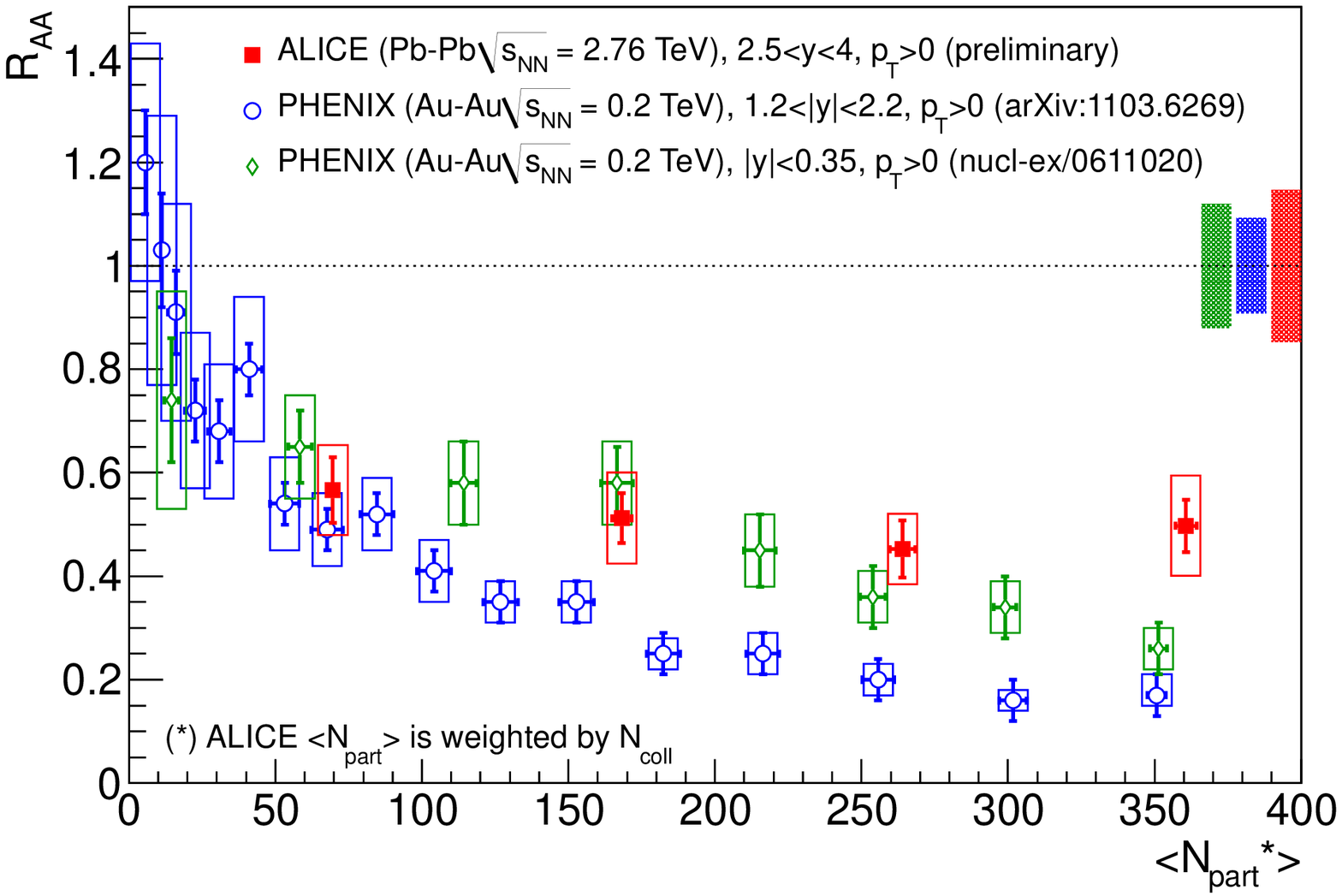}
   \caption{Left: Invariant mass distribution for opposite-sign muon pairs in the centrality class 0-10\% after mixed-event combinatorial-background subtraction. Right: J/$\psi$  $R_{\rm AA}$ as a function of $\langle N_{\rm part} \rangle$ in Pb-Pb collisions at $\sqrt{s_{\rm NN}}$=2.76 TeV compared to PHENIX results in Au-Au collisions at $\sqrt{s_{\rm NN}}$ = 200 GeV}
   \label{fig:JpsiInPbPb}
\end{figure}

Inclusive J/$\psi$ production was studied in Pb-Pb collisions at 2.76 TeV for both decay channels. The $R_{\rm AA}$ has been measured in the di-muon channel in centrality classes: 0-10\%, 10-20\%, 20-40\% and 40-80\%. The central to peripheral nuclear modification factor ($R_{\rm CP}$) was measured in the dimuon (dielectron) channel in 0-10\%, 10-20\% and 20-40\% (0-40\%), relative to the yield in the 40-80\% centrality class.
In the dimuon channel, the total analyzed integrated luminosity is 2.7 $\mu$b$^{-1}$ which corresponds to 17$\cdot$10$^6$ MB events. In Pb-Pb collisions, both tracks reconstructed in the muon tracking system are required to match reconstructed tracks in the muon trigger system. Different methods were applied to extract the J/$\psi$ signal, which is clearly seen in all centrality classes (Fig.\ref{fig:JpsiInPbPb}-left shows the invariant mass distribution after mixed-event background subtraction in the most central bin) \cite{Pillot11}. J/$\psi$ decays generated by MC simulations were embedded in real Pb-Pb events to study the performance of the ALICE muon spectrometer as a function of the collision centrality. 
A  4\% efficiency loss in the most central bin 0-10\% is observed, in agreement with the efficiency loss measurement based on the redundancy of the tracking chambers in each station. 
The J/$\psi$ nuclear modification factor (down to $p_{\rm T}$=0)  was evaluated (see Fig.\ref{fig:JpsiInPbPb}-right). The largest contributions to the $R_{\rm AA}$ systematic uncertainty are due to the signal extraction and the pp reference cross-section uncertainties. Fig 3, right panel, compares our results to those obtained by the PHENIX experiment \cite{PHENIX07}. We observe a weaker dependence with centrality  than that observed at RHIC, is measured. The $R_{\rm AA}(p_{\rm T}>0,~2.5<y<4)$ for the most central class is about a factor 2 larger than that measured by PHENIX with muons in the forward region; the difference is smaller, but still significant, when comparing to PHENIX at mid-rapidity. The $R_{\rm CP}$ was evaluated to make a comparison to the ATLAS results \cite{ATLAS11}. The ALICE $R_{\rm CP}(p_{\rm T}>0$,  $2.5<y<4)$ is smaller than the  $R_{CP}(p_{\rm T}>6.5{\rm ~ GeV}/c,~|y|<2.4)$ measured by ATLAS, thus the J/$\psi$ $R_{\rm AA}$  either increases with rapidity, or decreases with $p_{\rm T}$, or both. 
The measurement of the J/$\psi$ in the dielectron channel is challenging with the present statistics and large hadronic background.  However the signal has been extracted in the centrality class 0-40\% and $R_{\rm CP}$ with respect to the centrality class 40-80\% has been evaluated.  Within the large systematic uncertainties, the dielectron $R_{\rm CP}$ is compatible with ATLAS and ALICE di-muon $R_{\rm CP}$ measurements. Note that the ALICE J/$\psi$  measurement contains a contribution from B feed down. This contribution has been measured to be 10\% in pp collisions in our rapidity domain \cite{LHCb11} so the effect on the R$_{\rm AA}$ is expected to be small (at most a reduction in RAA of 10\%, if we assume binary scaling for bottom production).  Further analysis with the present data is being carried out: $p_{\rm T}$ and $y$ dependence of $R_{\rm AA}$, and narrower centrality bins for the most peripheral centrality classes.

\section{Conclusions}
The ALICE J/$\psi$ measurements in pp collisions at 2.76 and 7 TeV and in Pb-Pb collisions at $\sqrt{s_{\rm NN}}$=2.76 TeV have been presented.  The ALICE detector covers a large rapidity range down to $p_{\rm T}$=0. The J/$\psi$ yields, at $|y|<0.9$ and $2.5<y<4$, linearly increase with $|dN_{\rm ch}/d\eta|_{|\eta|<1.6}$, while the inclusive muon yield, arising largerly from from heavy-flavour decays, exhibits a stronger increase. ALICE has measured a uniform centrality dependence of the J/$\psi$ $R_{\rm AA}$ in Pb-Pb collisions, with magnitude larger than that measured at RHIC energies for the most central Au-Au collisions. The ALICE $R_{\rm CP}$ ($R_{\rm AA}$) is larger than that measured at higher $p_{\rm T}$ by the ATLAS (CMS) experiment at mid-rapidity. These results hint at J/$\psi$ regeneration in hot matter at LHC energies \cite{Andronic11}. Nevertheless, a number of effects related to interactions in cold nuclear matter remain unknown at LHC energies. 
For this reason a pPb run, which will address these nuclear effects, will be needed at LHC. Complementarily, most of CNM effects will cancel in the ratio of J/$\psi$ yield to that of open charm \cite{Dainese11} and a better sensitivity to hot nuclear matter effects on J/$\psi$ could be achieved.

\end{document}